\documentclass[hidelinks]{IEEEtran}
\usepackage{graphicx}
\usepackage{multirow, array, tabularx}

\newcommand{\letshpc}{\textit{Let's HPC}}

\IEEEoverridecommandlockouts

\begin{document}
%
\title{\letshpc{}: A web-based interactive platform to aid\\High Performance Computing education}

\author{\IEEEauthorblockN{Akshar Varma\IEEEauthorrefmark{2}, Yashwant Keswani\IEEEauthorrefmark{2}, Yashodhan Bhatnagar\IEEEauthorrefmark{2} and Bhaskar Chaudhury\IEEEauthorrefmark{1}
}
\\\IEEEauthorblockA{Group in Computational Science and HPC,
DA-IICT, Gandhinagar, India\\
\IEEEauthorrefmark{2}Contributed equally to this work,
\IEEEauthorrefmark{1}bhaskar\_chaudhury@daiict.ac.in
}}

\maketitle
\thispagestyle{plain}
\pagestyle{plain}

\begin{abstract}
\letshpc{} ({www.letshpc.org}) is an open-access online platform to supplement conventional classroom oriented High Performance Computing (HPC) and Parallel \& Distributed Computing (PDC) education. The web based platform provides online plotting and analysis tools which allow users to learn, evaluate, teach and see the performance of parallel algorithms from a system's viewpoint. The user can quantitatively compare and understand the importance of numerous deterministic as well as non-deterministic factors of both the software and the hardware that impact the performance of parallel programs. At the heart of this platform is a database archiving the performance and execution environment related data of standard parallel algorithms executed on different computing architectures using different programming environments; this data is contributed by various stakeholders in the HPC community. The plotting and analysis tools of our platform can be combined seamlessly with the database to aid self-learning, teaching, evaluation and discussion of different HPC related topics. Instructors of HPC/PDC related courses can use the platform's tools to illustrate the importance of proper analysis in understanding factors impacting performance, to encourage peer learning among students, as well as to allow students to prepare a standard lab/project report aiding the instructor in uniform evaluation. The platform's modular design enables easy inclusion of performance related data from contributors as well as addition of new features in the future.
\end{abstract}


\IEEEpeerreviewmaketitle

\section{Introduction}
"In theory, there is no difference between theory and practice. But, in practice, there is." - attributed to Jan L. A. van de Snepscheut and/or Yogi Berra. This is especially true of High Performance Computing (HPC)/Parallel and Distributed Computing (PDC) which combines aspects of both software and hardware to achieve the maximum possible performance out of a particular system for a particular problem. Classroom based teaching, focusing on theoretical aspects and book based concepts is not enough to fully convey the intricacies of HPC/PDC to students. Most of the efforts towards the development of curricula in PDC for undergraduate courses and fostering HPC education have been supplemented by educators in the form of sample lectures, tutorials, modules, books, software, recommended assignments, sample exercises, problem sets etc.~\cite{cderproject, CSinParallel,algorithms,quinn2}. These initiatives are primarily focused on the important areas of algorithm design, programming, computer architecture etc.,  and aid in introducing key parallelism concepts such as concurrency, dependency, tasks, threads, problem decomposition, data parallelism, recursion, synchronization, race conditions, resource sharing etc. to undergraduates~\cite{cderbook,hpcforscientist,tareador}. These efforts have significantly helped in popularizing HPC/PDC education at undergraduate level throughout the world, however more efforts are required to educate competent undergraduate students to learn parallel programming from the whole system's perspective. 

Theory of parallel algorithm design or theory of parallel computing is based on abstract concepts of time and memory which may ignore real life constraints of time and hardware resources for simplicity; and therefore do not take into account non-deterministic and hardware factors~\cite{algorithms}. The performance of a computer program depends on a wide range of factors like the nature of the algorithm, the machine (several hardware factors), the compiler, the runtime environment, the input, the measurement methodology etc. and their mutual interaction~\cite{parallelarchitecture, computerarchitecture, scientific-benchmark}. These are difficult aspects to impart from a pedagogic point of view using solely lectures and material from books~\cite{quinn1,pacheo}. Even if one perfectly understands the behaviour of a program and the properties of the targeted hardware system, one can not confidently understand and predict the behaviour of the program on the system without actually evaluating the program in the given software environment-hardware setting. For example, theoretical concepts like Amdahl's law put a linear limit on the speedup, while certain implementations can achieve superlinear speedups due to factors like optimal cache utilization~\cite{computerarchitecture, amdahls, amdahlsnew, gustafson}. 

Therefore, there is a clear gap between theoretical learning, implementation, and understanding/analysing the effect of non-deterministic factors on the performance of a program. Our web-based framework ({www.letshpc.org}) aims to address this gap using online plotting and analysis tools to better understand various aspects that determine the performance of programs in an HPC oriented setting~\cite{letshpc}. While our tool takes into account all the stakeholders in the HPC community, it is particularly useful for HPC course instructors and students.

Based on the experience of imparting courses on HPC and Parallel Programming at DAIICT, it has been reported that evaluation in HPC/Parallel Programming courses need to focus on lab assignments and projects as much as on conventional exams that test students' theoretical understanding~\cite{bhaskarc}. Lab assignments and projects are natural mechanisms for evaluation that allow students to analyse and realize the impact of various hardware/software factors on the performance of their own code. Our web-based platform has various tools that ease and streamline this process~\cite{letshpc}. 

The rest of this article is structured as follows: Section \ref{sect:design} describes the design philosophy of the \letshpc{} platform and details how it might be useful for various stakeholders in the HPC community. Section \ref{sect:website-overview} contains an overview of the website, the internal structure of the platform and the online tools. In Section \ref{sect:usage-example} we demonstrate how the platform can be used in the performance analysis of problems using the example of matrix multiplication. Section \ref{sect:future-conclusion} has concluding remarks including the current status and future plans for the \letshpc{} project. 

\section{Design philosophy of the \letshpc{} platform}\label{sect:design}

The design of the \letshpc{} platform has been made on the basis of two major guiding principles:
\begin{enumerate}
    \item Build web-based tools that are handy and useful to analyze the performance of a parallel program from a system's perspective and thereby help all stakeholders in the HPC/Parallel Programming community to understand the barriers to higher performance.
    \item Keep the design modular to allow easy storage, access, exchange and manipulation of data, and addition of more analysis tools without disturbing the platform.
\end{enumerate} 

Keeping in mind a need for a benchmark-like database for students to learn and for instructors to use to guide students, the central part of our platform is an archival database that contains all the necessary data from a given computational experiment to study performance on the basis of the software used (problem solving approach, serial and parallel code) and the execution environment used (machine, OS, compiler, parallel framework etc.) in the experiment.

Each of the tools interacts with the database and uses the data to construct plots that can be used to analyze performance. The tools have been designed in a way as to benefit instructors, students, self-learners and independent researchers in the learning and teaching of HPC. Used together these tools can make the study of HPC as smooth as possible, allowing students and instructors to focus on the important parts by removing mundane tasks such as generating plots and compiling a report of analysis.

\begin{figure}
\includegraphics[width=\linewidth, keepaspectratio]{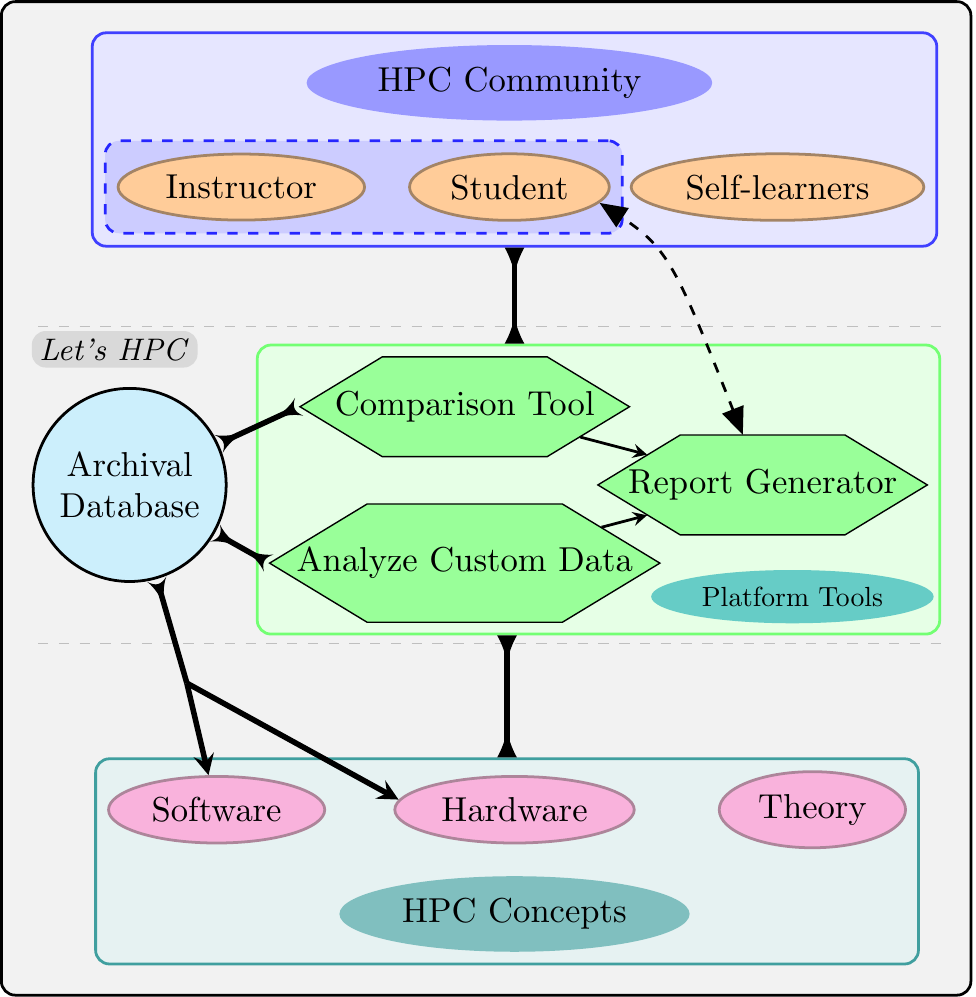}
\caption{Schematic of the conceptual design of the \letshpc{} platform illustrating how the components interact with stakeholders of the HPC education community (platform users) and with important HPC/parallel programming concepts.}\label{fig:design-philo}
\end{figure}

The four major components of the \letshpc{} platform, schematically shown in Figure~\ref{fig:design-philo}, are as follows:
\begin{itemize}
    \item The \textbf{archival database} containing software and hardware details of HPC experiments for various problems solved using various approaches on various machines. These act as a benchmark for the HPC community to study and teach from; analogous to the theory contained in textbooks, this will contain results and analysis which would throw light on the effect of various factors that are difficult to quantify in theory.
    \item A \textbf{comparison tool} with plotting features~\cite{google-charts} that allow users to analyze and compare the data from various benchmark experiments that have been contributed by members of the HPC community. This is useful when users want to analyze, compare, or learn from the performance that others have achieved using various implementation approaches on various machines and programming environments.
    \item \textbf{Analyze custom data:} a tool that allows anyone in the HPC community to analyze their own data. This is useful when users want to analyze results that they achieved using their own implementations-machine-environment combination using the plotting and analysis tools available on our platform.
    \item A \textbf{report generator} mainly aimed at students which allows them to upload their data and answer a basic list of well thought-out questions based on plots generated automatically. This allows for the generation of a uniform report that facilitates evaluation of labs/projects by instructors. It is also useful as a starting point for anyone who needs an HPC related report.
\end{itemize}

A brief review of the tools of our platform from the perspective of three major stakeholders (the instructor, the student, and the self-learner) in the HPC community that we perceive will benefit the most from this framework have been presented below:

\subsection{Course Instructors}\label{sect:instructors}
The primary mechanism for course instructors to use the \letshpc{} platform is via the archival database containing data from numerous problems and solutions approaches. These can be used to supplement the theoretical knowledge that students receive with analysis of performance of various approaches and machine factors. These would provide students with example analysis that they can easily grasp and replicate during their own work. The platform also contains a comparison tool which allows users to upload and compare their custom data with any of the existing data available in the database. This provides an additional means by which instructors can teach students; instructors can use this to showcase how certain factors can affect performance using real examples that would be difficult to showcase using theory and textbook knowledge alone. Our tools' focus on analysis ensures that students quickly realize the importance of all the factors that impact the overall performance. Section \ref{sect:usage-example} contains an example of how users could use the platform's archival database (and similarly the comparison tool) to study and analyse the performance of problems in an HPC setting.

The use of these two tools are analogous to how in more theoretical courses, textbooks can first be used to teach a concept and later examples can be solved to show how those concepts are used. Together these provide the means for instructors to provide students with an overall understanding of concepts in a manner similar to that available in more theoretical courses without losing the focus on the practical side of HPC.

Apart from bridging the theory-practice divide, one of the major difficulties faced by HPC course instructors is in the evaluation of students. HPC courses need to focus on practical aspects, in a manner which is different from a normal algorithm/programming course. HPC lab assignments are not as straightforward as other programming assignments; there is not necessarily any ``single correct answer''. From a pedagogical point of view, the focus is on understanding the reason for the performance achieved over merely achieving optimal performance. An HPC course needs to be lab/project oriented and in such cases instructors need a method to evaluate the students. A simple and common solution comes in the form of a report to be submitted by the students containing their understanding and analysis of the results. Our platform's report generator (see Section~\ref{sect:report}) is an apt tool for such cases. The systematic manner of the report generator in which questions have been divided into sections allows for the compilation of a proper, concise report which helps instructors to do a uniform evaluation.

\subsection{Students}\label{sect:students}

Students need to realize the importance of acquiring a holistic view of the system while learning parallel programming and they need resources via which they can learn how system factors beyond the code, such as the hardware and the programming environment play a role in performance. Generally students have access to only a couple of computing systems in the lab sessions of HPC courses, and even those may have similar configurations. Similarly, their lab assignments/projects may be restricted to merely focusing on getting correct answers and on the implementation approach due to a lack of resources as well as time. 
Our archival database provides students with an opportunity to analyse the performance of various HPC systems, from hardware differences to the effect of programming environment.
The platform also provides an opportunity to students to analyse and compare the performance of their implementations to those of others; this can happen at a course level or on a global level with students first learning from their peers and then from others in the HPC community. This will help students in easily learning more than one approach to parallelize a problem and also motivate them in improving the performance of their code.

From the point of view of students, making a polished report requires a lot of effort starting from measuring execution times, performing statistical preprocessing, generating plots, and finally performing various kinds of analyses to understand why a given performance is being achieved. Our platform's ``Report Generator'' tool automates much of this process for students. Apart from aiding in making reports our tool brings to notice many minor points of analysis that students may miss out on. This ability to look at the whole system and being consciously aware of the fact that each component can impact performance can help in writing better codes, more attuned to the system.

\subsection{Self-learners}\label{sect:self-learners}
Assuming willing users with access to decent computational resources, our framework can be used by anyone to study HPC on their own. Our platform supplements theoretical, book-based knowledge very well and can act as a ``textbook'' to study non-deterministic factors. The available tools aid self-learners to get rid of mundane tasks and focus on understanding HPC from the whole system's perspective. 

Apart from these stakeholders, independent researchers can also find use for the platform as a repository for benchmarked data, to contribute their own data and as a forum for discussion with other HPC educators/researchers.

\section{Overview of the website and tools}\label{sect:website-overview}
\letshpc{} has been developed using the standard MEAN stack web framework~\cite{mean-stack}. The MEAN (MongoDB, ExpressJS, Angular, Node) stack allows for modularity in the structure of the whole platform from backend to frontend. Each piece works independently, with Node-MongoDB-Express forming the server side framework to serve database requests and Angular forming the client side scripting framework. Being a popular choice for many developer communities, the open source technologies used in the MEAN stack are continuously maintained and upgraded by their respective communities. This also implies that regular upgrading and addition of new features to the platform is fairly straightforward. 

The plotting and analysis tools all use results present in the database depending on the user's access permissions\footnote{Currently, the users module is under construction and therefore, only the public data has been made available.}. \letshpc{}'s archival database is \textit{conceptually} structured into different databases based on access permissions of users. The conceptual division based on access permissions are:

\begin{itemize}
    \item Public benchmark data accessible to everyone, even without an account. These would contain data that has been curated and made public by various contributors.
    \item Data of courses accessible to instructors of registered institutes. Each batch of students may have (conceptually) different databases and the instructor would have access to and control over all of this data, with the ability to make data of a given assignment public (to other students) as and when they want to.
    \item Data of students registered in courses offered by instructors. Students can directly upload their data and analyze it using the tools provided. Instructors will have access to this data and the analysis, for evaluation.
    \item Private data of individual contributors, only accessible to them. This will primarily be data pertaining to unpublished, ongoing research; collaboration among contributors would also be possible in this setting.
\end{itemize}

In Section~\ref{sect:data-collection} we provide details of the data collection methodology that we have used to collect the starting data for our database. Section~\ref{sect:data-filtering} illustrates how we set filters for accessing data from the database. Section~\ref{sect:report} discusses the report generator tool.

\subsection{Data Collection Methodology} \label{sect:data-collection}
\begin{table}
\renewcommand{\arraystretch}{1.25}
\centering
\caption{Illustrative set of categories and sample problems}\label{tab:problem-category}
\begin{tabular}{|c|l|}
\hline
  \textbf{Category}  & \multicolumn{1}{c|}{\textbf{Problems}} \\
  \hline
\multirow{2}{*}{Linear Algebra}  & Vector Dot Product\\
& Matrix Multiplication\\
\hline
\multirow{3}{*}{Reduction, Scan, Sort}  & Array Sum\\
& Scan\\
& Quicksort\\
\hline
\multirow{2}{*}{Image Processing}  & Grayscale conversion\\
& Median filtering\\
\hline
\multirow{2}{*}{Divide and Conquer}  & Monte Carlo\\
& Pi using Series Sum\\
\hline
\end{tabular}
\end{table}

At present, data collected from various shared memory systems using the OpenMP environment is being made public on the platform. For each machine, serial and parallel versions of the code are run multiple times for a range of problem sizes and data of execution times is collected. Two kinds of time data have been collected:
\begin{itemize}
    \item Algorithmic (ALG) time: This includes only the time for the core algorithm (memory access and operations), and does not include any time for the I/O part.
    \item End-to-end (E2E) time: This counts the entire run-time of the program. This includes the time taken for I/O (reading the input test-case files and writing the output data files as well) in addition to ALG time.
\end{itemize}

Apart from serial and parallel codes, we collect meta data containing the broad problem category and problem (examples in Table~\ref{tab:problem-category}), the algorithmic description (implementation approach), processor related machine details, and programming environment details (OS, compiler, OpenMP versions). All this meta data along with the problem size and number of processors/threads used in a particular execution of the program defines what we call a ``configuration''. Each configuration has been run 10 times to minimize anomalies which might arise due to the non-deterministic hardware dependencies (other processes, OS scheduling, etc.). Before plotting any of the data, our tools perform statistical preprocessing as necessary\footnote{Based on suggestions in \cite{scientific-benchmark}, done whenever data available.  } and then plot the execution time, the relative speedup\footnote{The relative speedup $S=T_p/T_s$, where $T_p$ is the time for parallel code and $T_s$ is the time for serial code using the same approach.}, the efficiency, and the Karp-Flatt metric.

The code template used to calculate the execution times in codes can be found on the website. The hardware specifications and OS, compiler versions are also collected. In our case, we used the following commands:

{\footnotesize
\begin{itemize}
    \item \texttt{cat /proc/cpuinfo} \hfill (CPU)
    \item \texttt{lscpu} \hfill (CPU)
    \item \texttt{uname -a} \hfill (OS)
    \item \texttt{gcc --version} \hfill (Compiler)
\end{itemize}
}

As discussed earlier, any HPC system/model consists of 3 important blocks, namely the system hardware, the algorithm (specific implementation) and the programming environment. The data we have collected combine to provide a holistic picture of any such HPC system and allow users to perform analysis from the whole system's perspective rather as we illustrate in Section~\ref{sect:usage-example}.

Each of these blocks and their component parts have to be analyzed by segregating relevant components which affect the performance, and studying each of them as well as the correlation among them. We now discuss how this segregation has been naturally designed into our platform.

\subsection{Data Filtering Process}\label{sect:data-filtering}
\begin{figure}
\begin{center}
\includegraphics[width=0.45\textwidth, keepaspectratio]{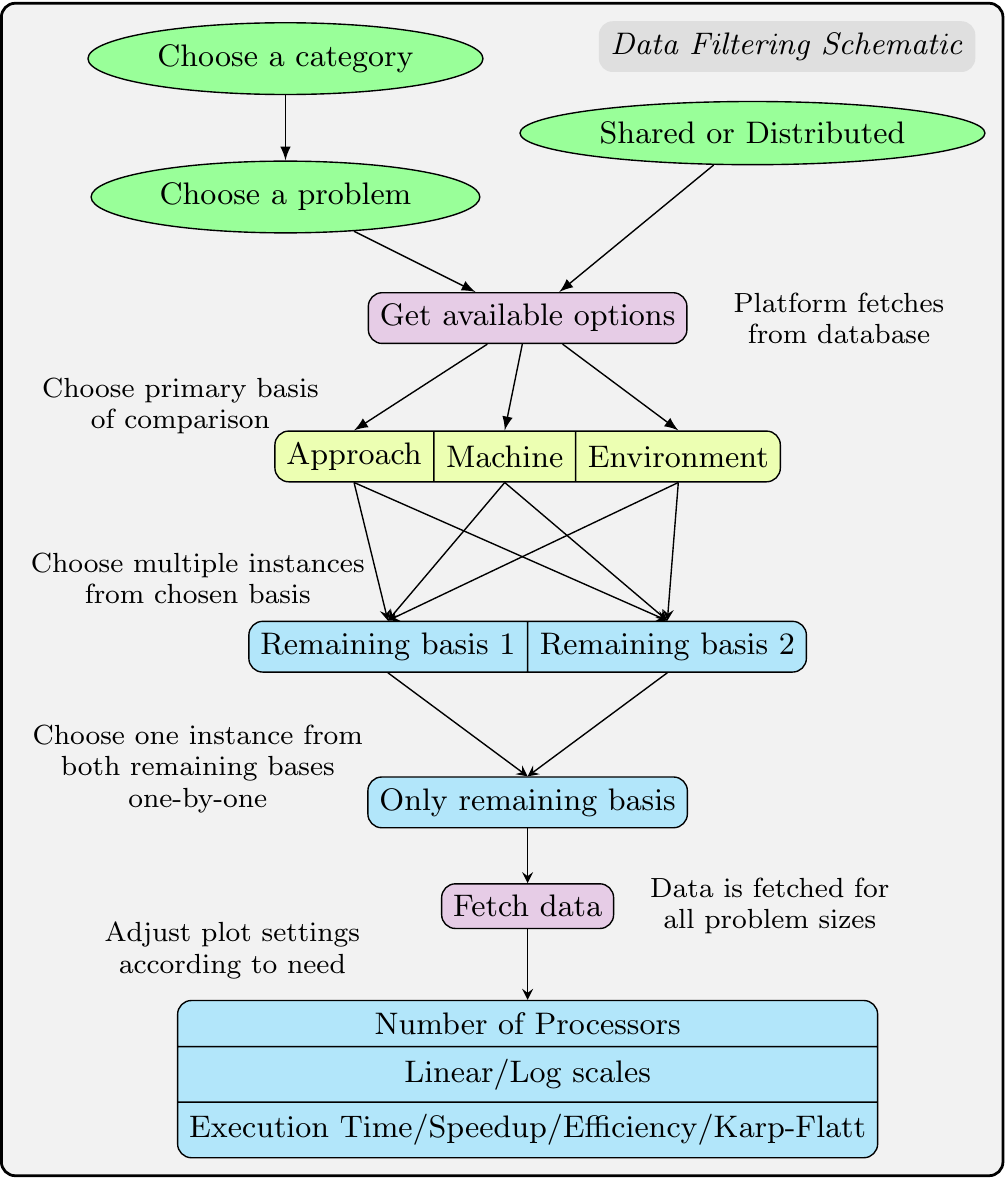}
\end{center}
\caption{Schematic explaining how the data filtering process works in the \letshpc{} platform.}\label{fig:data-filter}
\end{figure}

Our platform's plotting and analysis tools get data for comparison from the database using a flexible filtering process that can be expanded with additional features and provides opportunities to compare a wide variety of factors that make up the HPC system. Figure~\ref{fig:data-filter} contains a conceptual schematic that describes the flow of the tool when users filter data according to their requirements.

The user first selects a category of problems and a problem belonging to that category. Along with this, shared or distributed memory system is selected. For the example in Section~\ref{sect:usage-example}, we would select ``Linear Algebra'' and ``Matrix Multiplication'' along with ``Shared memory system''.

The website then fetches data from the database and provides three options (compare either approaches, or machines or programming environments) for the primary basis of comparison. The user makes a choice here; in the example in Section~\ref{sect:usage-example} we illustrate using both ``Approaches'' and ``Machines''. Whichever basis is chosen, multiple instances can be chosen which the user wants to compare. The platform then provides a list of the remaining options in each of the remaining bases. The user selects one instance from each and data is fetched for all the problem sizes. The user can then proceed to adjust the plot settings according to their preference and study the plots. Users can also download the plots and include them in the report generator.

\subsection{Report Generator}\label{sect:report}
\newcolumntype{V}{>{\centering\arraybackslash} m{.4\linewidth} }
\renewcommand{\arraystretch}{1.25}
\begin{table}
\centering
\caption{Sample questions for the Report Generator}\label{tab:report-short}
\begin{tabular*}{\linewidth}{@{\extracolsep{\fill}}|c|p{0.55\linewidth}|}

\hline
  \textbf{Section}  & \multicolumn{1}{c|}{\textbf{Question description}} \\
  \hline
\multirow{2}{*}{{Basic Description}}  & Basic questions asking for description of the Serial and Parallel approaches used.\\
\hline
\multirow{3}{*}{{Complexity, analysis}} &  Questions regarding complexity of the implementations, memory accesses, computations, theoretical speedup, etc. \\
\hline
\multirow{2}{*}{{Curve based analysis}} &  Analysis of Execution time, Speedup, Efficiency and Karp-Flatt metric plots.\\
\hline
  \multirow{3}{*}{{Further detailed analysis}} &  Detailed analysis on the basis of concepts like cache coherence, false sharing, granularity, load balancing, etc.   \\
  \hline
  \multirow{4}{*}{{Additional analysis}}  &  Analysis of various miscellaneous factors that impact performance; advantages/disadvantages and difficulties with respect to the implementation. \\
\hline
\end{tabular*}
\end{table}

As we have already discussed in Section~\ref{sect:students}, students need a mechanism to quickly and efficiently make reports based on their analysis that instructors can evaluate uniformly and easily. The report generator part of our platform has questions designed to significantly ease this process for students. It allows students to upload their results, get plots automatically generated and a set of questions which students answer on the basis of their analysis. They get a \texttt{.tex} file along with the required plots that they can then modify as required and compile to get a PDF report. It forms the last step in the chain of any HPC analysis, and our platform automates all aspects starting with students uploading their execution results and ending with students getting a report containing their analysis. Table~\ref{tab:report-short} contains the broad idea of the division of questions into various sections and a short description of the kind of questions.

\section{Performance Analysis using Platform Tools}\label{sect:usage-example}
\begin{figure}
\begin{center}
\includegraphics[width=\linewidth, keepaspectratio]{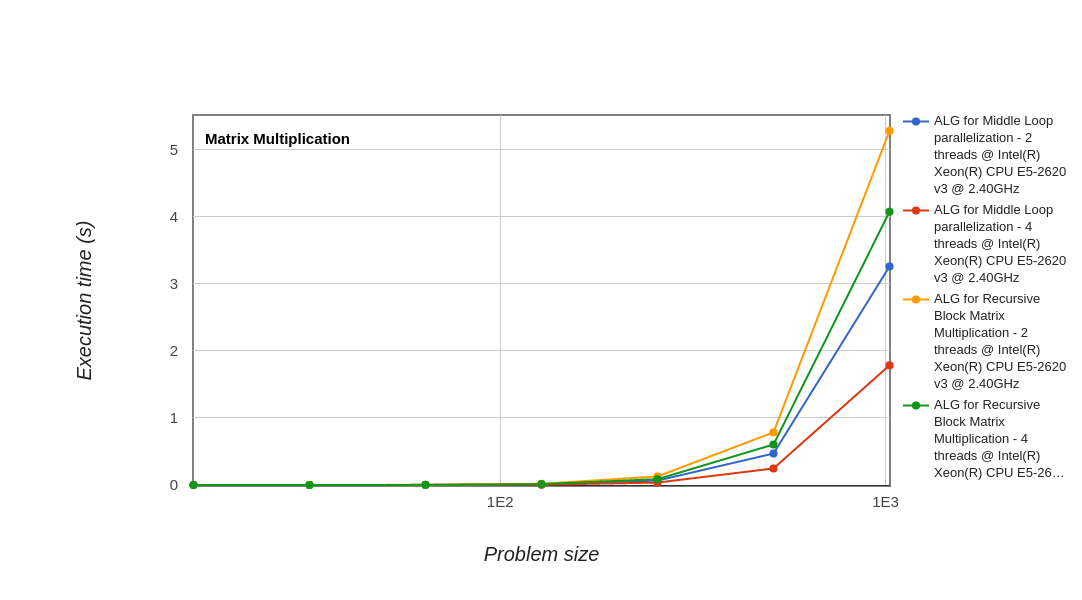}
\includegraphics[width=\linewidth, keepaspectratio]{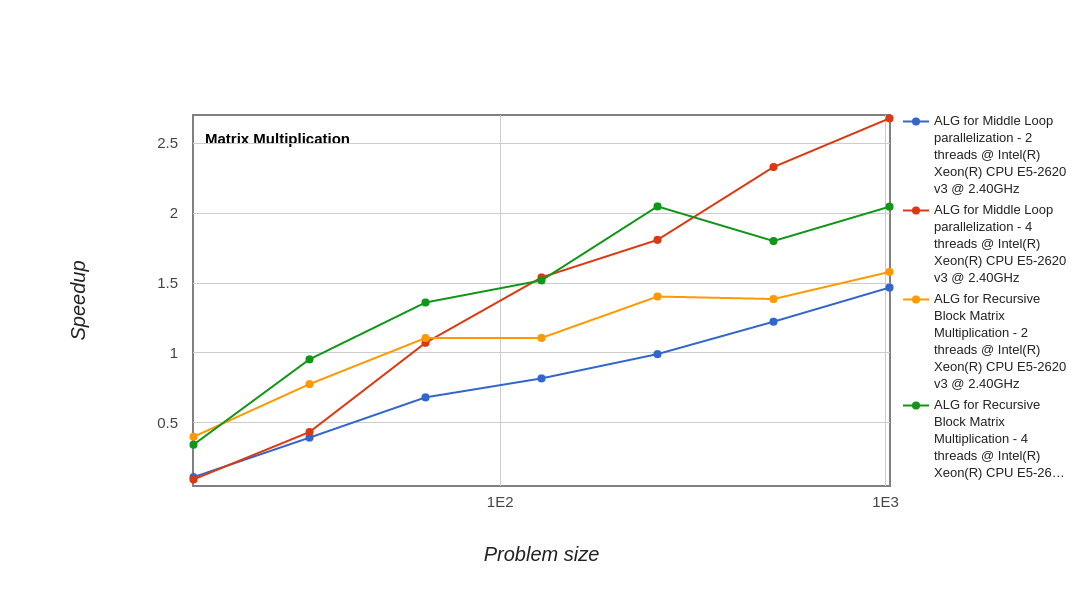}\\
\includegraphics[width=\linewidth, keepaspectratio]{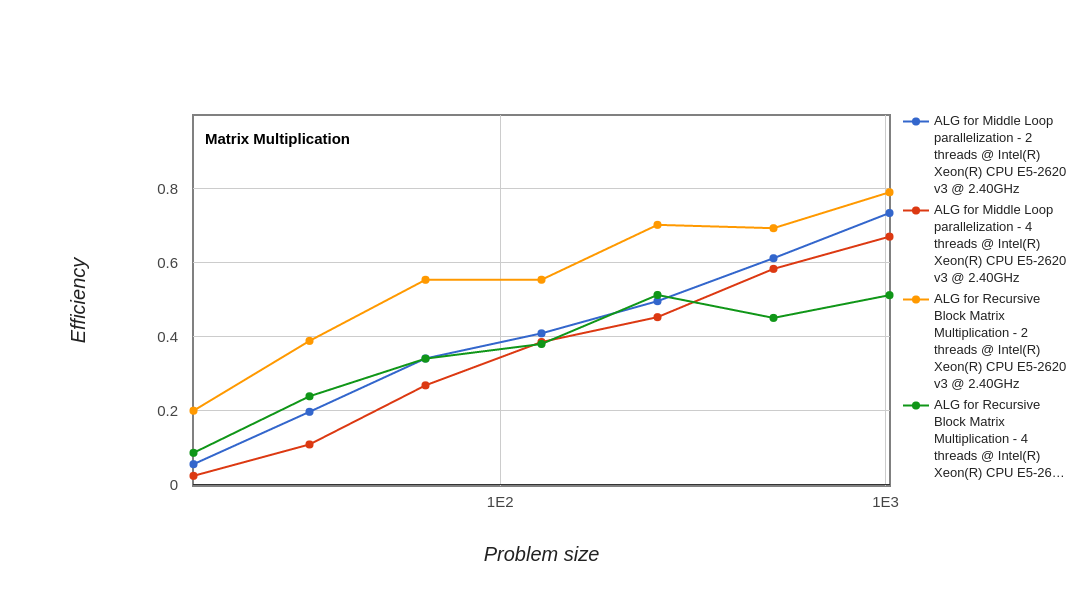}\\
\includegraphics[width=\linewidth, keepaspectratio]{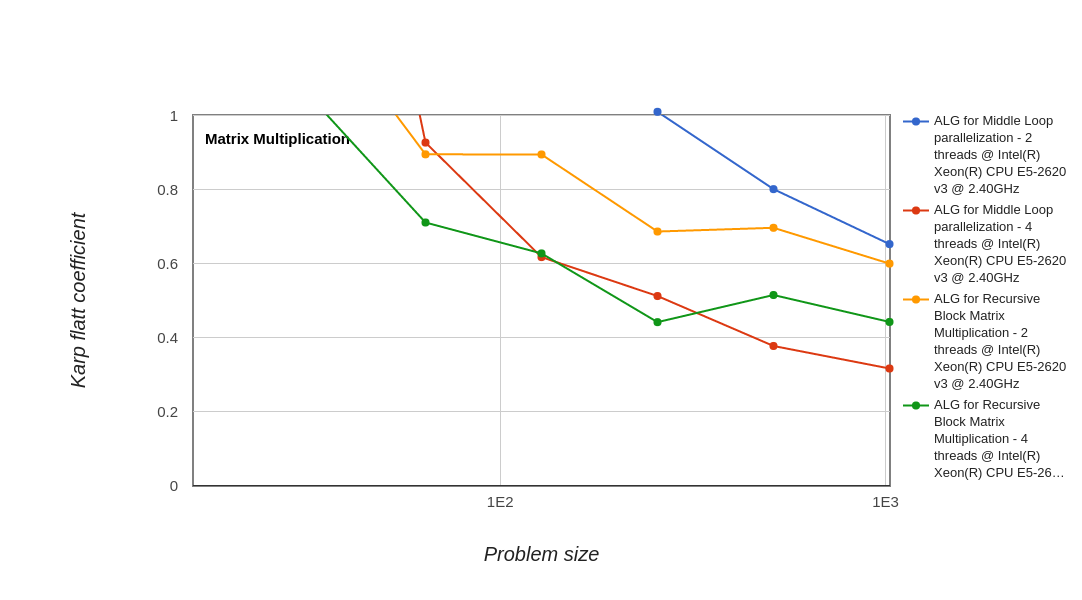}\end{center}
\caption{Plots generated using the \letshpc{} tool (www.letshpc.org) for comparing the two approaches of matrix multiplication.}\label{fig:approach-plots}
\end{figure}

In this section we show how the tool can be used to compare the performance given by an HPC system using some standard laws, ground rules and guidelines for interpretation of results~\cite{scientific-benchmark, karpflatt, grama}. There are numerous approaches possible for Matrix multiplication; we illustrate the usage of the tools of the \letshpc{} platform by comparing two approaches and two machines.

\subsection{Comparing Approaches}
We compare recursive block multiplication approach (Approach 1) and parallelization of the middle loop using \texttt{`\#pragma omp parallel for'} approach (Approach 2). We plot the curves for the parallel version for 4 threads and 2 threads and then analyze those curves to understand the performance of the two approaches. We analyze averaged results (Figure~\ref{fig:approach-plots}) of runs on an Intel\textsuperscript{\textregistered} Xeon\textsuperscript{\textregistered} E5-2620 processor at 2.4 GHz.

\subsubsection{Analysis based on the time graph}
We can clearly see (comparing the times for 4 threads) that Approach 2 is better than Approach 1 in terms of execution time. The number of recursive calls in the recursive block method is a function of the problem size, so as the problem size increases, more number of recursive calls are made, and hence the time is much greater than that for middle loop parallelization approach.

\subsubsection{Analysis based on the (relative) speedup graph}
The first thing we notice is that relative speedup only occurs for problem size $n>=64$ (multiplication of two 64 * 64 matrices) for both the approaches. This is because for smaller problem sizes, the parallelization overhead in terms of initialization of threads and scheduling is significant in comparison to the parallelization achieved. For larger problem sizes, the computational part of the code is significantly higher than the overhead and we notice a speedup.

In the recursive block approach, we see that the speedup tends to saturate at around 2. Knowing that the recursive block approach has been implemented using recursive calls suggests that the increasing number of recursive calls may be causing this saturation of speedup but this is unlikely to be the major cause since the speedup is relative to the same serial approach. More likely reasons, caused by the recursive implementation, are scheduling overheads and memory access overheads that are incurred due to the lack of any discernible pattern in memory accesses.

In comparison, the steady slope of the speedup curve for the middle loop parallelization approach show that it is a better approach in terms of scalability as the problem size increases. This is as expected since this approach has a much better memory access pattern compared to the recursive calls of the other approach and hence allows for better utilization of the cache memory. We note that this approach provides better results in terms of speedup, but that should not be the final answer and we continue our analysis to further understand the reasons for the achieved performance.

\subsubsection{Analysis based on efficiency}
The efficiency of the middle loop approach increases roughly linearly as the problem size increases. This is because as the problem size increases, the overhead increases as $O(n)$ but the computations increase as $O(n^3)$, thus causing the computation part to dominate over the overhead and thereby causing an increase in efficiency as the problem size increases.

For the middle loop approach, looking at a cross section ($x=c$ line) of the efficiency plot with multiple threads plotted, the efficiency is greater for 4 threads as compared to 2 threads. This suggests that this approach is scalable as the number of processors is increased.

A similar analysis for the recursive block based approach shows that the efficiency for 2 threads is more than that for 4 threads, which indicates that the problem does not scale as we increase the number of processors. This is similar to the lack of scalability with respect to problem size.

\subsubsection{Analysis based on Karp Flatt}
We use the graphs of the Karp-Flatt metric to corroborate the analysis we performed using the other plots. These plots represent the experimentally determined serial fraction of the code. For the recursive block approach, we see that the serial fraction starts to saturate towards the higher problem sizes, while for the middle loop approach the serial fraction continues to decrease. This provides further evidence for our earlier analysis.

In the case of processors, a cross section approach doesn't provide striking evidence but does seem to match the analysis that we did using the efficiency curves. For recursive block approach, the serial fraction seems to be slowly saturating while for the middle loop approach the serial fraction shows a slow but steady decline.

Overall, using all these plots, one can analyze the two approaches for solving the matrix multiplication problem and conclude that the middle loop parallelization approach is better and more scalable than the recursive block matrix approach. Both the approaches suffer from parallelization overheads, but the middle loop approach is comparatively better from a parallelism point of view.

\subsection{Comparing Machines}
\begin{figure}
\begin{center}
\includegraphics[width=\linewidth, keepaspectratio]{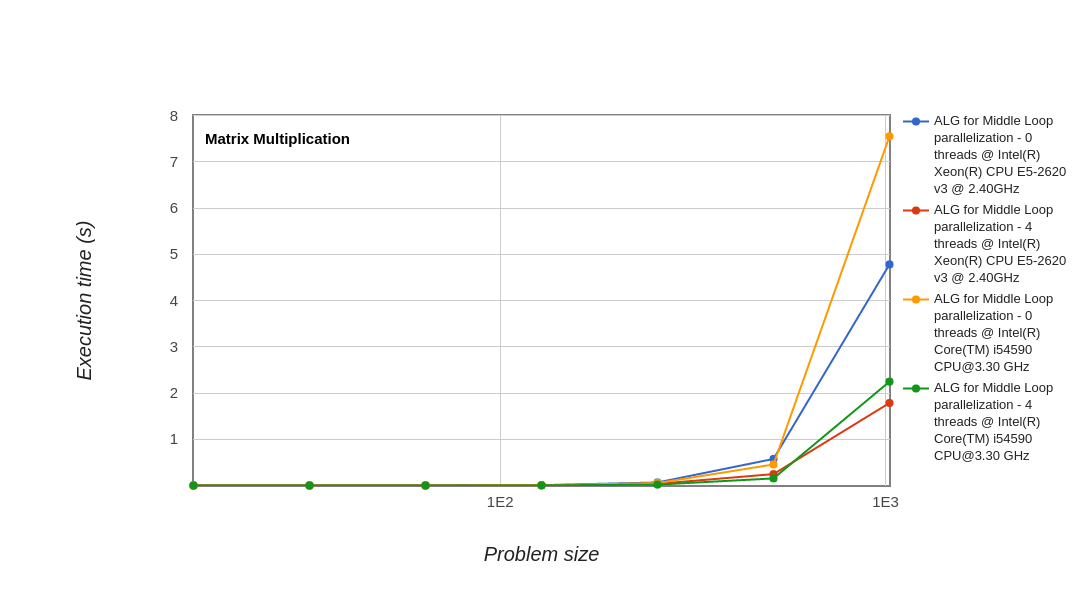}
\end{center}
\caption{The plot generated using the \letshpc{} tool when comparing the two machines.}\label{fig:machine-plots}
\end{figure}

A similar analysis is possible while comparing machines but here we focus only on the machine specific aspects that our tool helps in isolating. We compare an Intel\textsuperscript{\textregistered} Xeon\textsuperscript{\textregistered} E5-2620 processor at 2.4 GHz (Machine 1) with an Intel\textsuperscript{\textregistered} Core\textsuperscript{\texttrademark} i5-4590 processor at 3.3 GHz (Machine 2).

From the execution times (Figure~\ref{fig:machine-plots}), we see that the time of the serial code for Machine 1 is almost 1.5 times less than that for Machine 2. It would be a simple but mistaken assumption to think that since Machine 2 has a processor with a greater clock speed, it would tend to perform better. Our platform provides the basic specifications of the processors and a link to the full specification by the respective vendors. In this case, since there are a lot of memory accesses taking place in the approach, the cache sizes also start to matter. Both the machines have similar L1 and L2 caches, however the L3 cache for Machine 1 is 20 MB and for Machine 2 is 6 MB. Further, looking at the full specifications, we see that the maximum memory bandwidth of the Integrated Memory Controller for Machine 1 (59 GB/s) is more than twice that for Machine 2 (25.6 GB/s) which also explains the difference in times.

Thus, the plotting tools of our platform allow for a holistic analysis between approaches or machines (or programming environments) which enable users to completely understand the system.

\section{Future Scope and Concluding Remarks}\label{sect:future-conclusion}
Our framework is designed in a highly modular manner, providing numerous features that are useful to various stakeholders in the HPC community. The archival database is apt for acting as a benchmark for the HPC community particularly from a pedagogical perspective. At the same time, it is a tool that acts as a record of the HPC capabilities of various machines across years and can become a resource later for studying the gradual evolution of multicore architecture and performance of parallel algorithms on such architecture. The plotting and analysis tools built on top of it, including the comparison tool and the tool allowing analysis of custom data both fill a gap in current approaches to HPC education. Our platform supplements theoretical and algorithm oriented education by showcasing the importance of keeping a holistic, system's perspective to the study and analysis of performance of HPC systems. The report generator tool makes HPC education easier for both instructors and students by bringing the focus to concepts that matter and allowing automation of trivial, mundane tasks.

The modular design of our platform allows for numerous, easy ways to improve the framework and to add features in the future. We are currently in the process of developing discussion forums allowing comments for each approach-machine-environment configuration in the database. Currently, the involvement of the HPC community on this platform is a passive involvement with most users only studying the data provided. The introduction of discussion forums which will allow users to participate more interactively with other users. Extending this with HPC courses in mind, we will also introduce more refined forum abilities for HPC courses, with peer learning oriented features for students and teaching and evaluation related features for instructors. This would not only bring in very fine access permissions which can enable all course assignments and projects to be done and evaluated but also facilitate active collaboration. Further features along with discussion forums can make the whole process of collaboration more streamlined.

We plan to include some advanced tools and models in the existing framework which will offer additional insights to educators and students on how to improve performance~\cite{roofline}. One such improvement is extending the report generator tool to allow users (especially instructors) to make custom report generators. As a long term goal, we envision the \letshpc{} project to become a platform satisfying all requirements for HPC education, providing additional tools that combine to give an end-to-end solution starting by adding scripts to automate data collection and statistical tools allowing analysis of variability of performance data. Eventually, this will include building tools for online comparison of experiments run on various heterogeneous architectures and accelerators as well as different emerging cloud computing paradigms, particularly for computationally expensive scientific algorithms~\cite{cloudhpc1}. As a more generic service, we want to eventually allow users to upload just their codes and then automate the whole end-to-end process using servers to run codes and allow users to directly move to analyzing performance results. This would truly allow HPC to be adopted by those without high end HPC resources.

\section*{Acknowledgments}
The work has been carried out using the HPC resources at DA-IICT. The author B. Chaudhury would like to acknowledge the support received from NSF through NSF/TCPP CDER Center Early Adopter Award Program. We would like to thank students of CS301 High Performance Computing course (2016-2017) at DA-IICT for providing some of the initial serial and parallel codes on which this tool was tested.

\bibliographystyle{unsrt}
\bibliography{EduPar17-letshpc}


\end{document}